\documentclass[aps,prd,showpacs,showkeys,reprint,longbibliography,amsmath]{revtex4-1}
\usepackage{latexsym,graphicx,subfigure,verbatim}

\begin{document}

\title{Quantum cosmology in Ho\v{r}ava-Lifshitz gravity}
\date{\today}
\author{O. Obreg\'on}
\email{octavio@fisica.ugto.mx}
\author{J. A. Preciado}
\email{preciado@fisica.ugto.mx}
\affiliation{Departamento de F\'{\i}sica, Divisi\'{o}n de Ciencias e Ingenier\'{\i}as, Campus Le\'{o}n, Universidad de Guanajuato, Loma del Bosque No. 103 Colonia Lomas del Campestre, C.P. 37150, Le\'{o}n, Guanajuato, M\'{e}xico.}

\begin{abstract}
  Quantum cosmology is studied within the framework of the minimal quantum gravity theory proposed by Ho\v{r}ava. For this purpose we choose the Kantowski-Sachs (KS) model and construct the corresponding Wheeler-DeWitt equation. We study the solution to this equation in the ultraviolet limit for different values of the running parameter $\lambda$ of the theory. It is observed that the wave packet for this Universe changes completely compared with the one observed in the infrared (general relativity) regime. We also look at the classical solutions by means of a WKB semiclassical approximation. It is observed that if $\lambda$ takes its relativistic value $\lambda = 1$ a generalized KS metric is obtained which differs from the usual KS solution in general relativity by an additional term arising from the higher-order curvature terms in the action and which dominates the behavior of the solution for very small values of the time parameter. We discuss the physical properties of this solution by comparing it with the usual KS solution in general relativity. The resulting solution has no horizons but singularities.
\end{abstract}


\maketitle

\section{Introduction}
As it is well-known, the gravitational field is the only interaction that has not been successfully fitted into the general framework of quantum theory. The theory describing this interaction, to the best of our current knowledge, is Einstein's general theory of relativity (GR). The experimental tests passed by GR are impressive, however, it has its own limitations in a more fundamental level. For instance, it suffers from notorious divergence problems and an ill ultraviolet (UV) behavior due to its nonrenormalizability. These issues suggest that GR is a purely low-energy or large-distance approximation to some more fundamental underlying model: a quantum theory of gravity. A good formulation of such a theory would be expected to solve the problems mentioned above, having GR as its classical limit.

In this context, Petr Ho\v{r}ava proposed an interesting approach whose formulation seems to provide a better starting point in the search to understand quantum gravity \cite{Horava1,Horava2}. The central idea of his theory is to combine gravity with the concept of anisotropic scaling between space and time, motivated by the recent developments in the study of condensed matter systems. Based on this principle, higher spatial-derivative correction terms may be added to the standard Einstein-Hilbert action such as different powers of the spatial curvature. This improves the UV behavior of the graviton propagator but at the cost of giving up the Lorentz invariance as a fundamental symmetry of the theory. Instead of that it is expected to emerge as an accidental symmetry in the infrared (IR) regime where general covariance must be restored.

In the minimal version of the theory the nature of the modifications is governed by the gravitational analog of the ``detailed balance'' principle, frequently used in the study of the dynamics of nonequilibrium systems, and the so-called ``projectability condition'' which restricts the lapse function to be a function of time only \cite{Horava2}. However, it is possible to generalize this formulation by relaxing any or both of these conditions. This has lead to the projectable and nonprojectable versions of the theory \cite{Visser1,Visser2,Visser3,Blas1,Blas2}.

It has been claimed that Ho\v{r}ava's first proposal and some of its extensions suffer from several issues related to a badly behaved scalar mode \cite{Niz,Blas3,Cai1,Koyama1,Papazoglou1}. However, it has desirable features which make it worth exploring. Many aspects of the theory have been discussed in the literature, particularly, cosmological and black hole solutions have been obtained \cite{Kiritsis,Calcagni,Pope,Cai2}. Many issues of cosmology arising from it have also been analyzed (see, for example \cite{Mukohyama}). In this context, models which could provide information about the quantum properties of this theory would be of interest. For this purpose, the simplest models one can build are those depending only on time: quantum cosmological models. We choose the Kantowski-Sachs (KS) model \cite{Kantowski} as it has been extensively studied in the framework of GR where it is associated through a well-known diffeomorphism with the Schwarzschild and Schwarzschild (anti)-de Sitter black holes \cite{Hawking,Zacarias1,Zacarias2}. A Wheeler-DeWitt (WDW) equation for this model is derived in the context of the minimal theory proposed by Ho\v{r}ava. We study the solutions to this equation in the UV regime for different values of the parameter $\lambda$ of the theory, comparing them with the corresponding solutions derived in the infrared regime (GR) by constructing the respective Gaussian-weighted wave packets in these two limits. Even though $\lambda$ is a free parameter susceptible to quantum corrections \cite{Horava2}, a particularly interesting result is obtained when $\lambda$ takes its relativistic value $\lambda = 1$, where the minisuperspace variables exchange their role in this limit compared with their usual GR behavior. This UV behavior is similar for other values of $\lambda$ too. This suggests that for other quantum models, in the realm of Ho\v{r}ava-Lifshitz gravity, not only those related with quantum cosmology, we should expect a very different behavior in the UV limit compared with the one in the IR regime (GR). We also look at the classical solutions by means of a WKB method. It is observed that for the specific case $\lambda = 1$, a generalized KS metric is obtained which differs from the usual KS solution in GR by an additional term arising from the higher-order spatial curvature terms in the action and which dominates the behavior for very small values of the time parameter as expected. We show the behavior of this generalized metric by comparing it with the one in Einstein's gravity.

The organization of this paper is as follows. In Sec. \ref{Theory} we review the general features of gravity models with anisotropic scaling and their Hamiltonian formulation. In Sec. \ref{KSGR} we briefly review the minisuperspace framework of quantum cosmology. Here the well-known WDW equation for the KS minisuperspace model is presented, and the WKB method is used to obtain the corresponding classical solution in GR. In Sec. \ref{KSHL} we derive the WDW equation for this model within the framework of the minimal theory proposed by Ho\v{r}ava. In particular, the UV regime is analyzed and it is shown that the behavior of the wave packet is completely different compared with the one observed in GR. Then a WKB approach is also performed to the generalized WDW equation and the corresponding classical equation is analytically solved for the case $\lambda = 1$ in order to obtain a generalized KS metric. We discuss its general physical properties in comparison with those of the standard KS solution of GR. Finally Sec. \ref{Conclusions} is devoted to discussion and conclusions.

\section{The Theory}\label{Theory}
\subsection{The Action}\label{TheAction}
In gravity theories with anisotropic scaling the degree of anisotropy between the space and time coordinates is characterized by a dynamical critical exponent $z$ \cite{Horava2}, such that
\begin{equation}\label{AS:z}
  \mathbf{x}\rightarrow b\mathbf{x},\qquad t\rightarrow b^{z}t.
\end{equation}
The gauge symmetries are those spacetime diffeomorphisms that preserve a preferred foliation $\mathcal{F}$ of the spacetime manifold $\mathcal{M}$  by fixed time slices, known as foliation-preserving diffeomorphisms $\text{Diff}_{\mathcal{F}}(\mathcal{M})$, generated by the infinitesimal transformations
\begin{equation}\label{symmetries}
  \delta x^{i} = \zeta^{i}(t,\mathbf{x}), \qquad \delta t = f(t).
\end{equation}
Thus, the spacetime manifold is equipped with a causal structure compatible with the preferred role of time implied by the anisotropic scaling \eqref{AS:z}. In this context, it results convenient to consider the Arnowitt-Deser-Misner (ADM) decomposition of spacetime and to construct the action in terms of the spatial metric $g_{ij}$, the shift vector $N_{i}$ and the lapse function $N$ \cite{ADM}. With all these ingredients the most general action of this class of gravity models in $D+1$ dimensions takes the form \cite{Horava3}
\begin{equation}\label{Action:AS}
  S = \frac{2}{\kappa^{2}} \int dtd^{D}\mathbf{x} \sqrt{g}N \left(K_{ij}K^{ij} - \lambda K^{2} - \mathcal{V} \right),
\end{equation}
where $\kappa$ and $\lambda$ are coupling constants, and $K_{ij}$ is the extrinsic curvature tensor of the preferred time foliation defined by
\begin{equation}\label{Kij}
  K_{ij} \equiv \frac{1}{2N} (\dot{g}_{ij} - \nabla_{i}N_{j} - \nabla_{j}N_{i}).
\end{equation}
The first two terms in the action \eqref{Action:AS} represent the most general kinetic term invariant under $\text{Diff}_{\mathcal{F}}(\mathcal{M})$, and $\mathcal{V}$ is an arbitrary potential term built on $g_{ij}$ and its spatial derivatives including all those terms compatible with $\text{Diff}_{\mathcal{F}}(\mathcal{M})$ and depending on the desired value of $z$. Originally $z = D$ was chosen by power counting arguments ensuring a dimensionless coupling constant $\kappa$ pursuing renormalizability \cite{Horava2}. As a first attempt, Ho\v{r}ava restricted $\mathcal{V}$ to satisfy the so-called ``detailed balance'' and projectability conditions. The former limiting the number of independent terms by constructing $\mathcal{V}$ as the square of the equations of motion of an action in one lower dimension, and the latter restricting the lapse field to be a function of time only \cite{Horava1,Horava2}. There seems to be no physical arguments behind these conditions, however, they allow a mathematically consistent formulation of gravity with anisotropic scaling preserving the essential properties of this new class of gravity models.

Then the action of this nonrelativistic theory, under all these considerations, in $3+1$ dimensions is given by
\begin{align}\label{HoravaAction}
  \nonumber S= &\int dtd^{3}\mathbf{x}\sqrt{g}N \bigg\{ \frac{2}{\kappa^{2}}(K_{ij}K^{ij} - \lambda K^{2}) - \frac{\kappa^{2}}{2w^{4}}C_{ij}C^{ij}
  \\ \nonumber & + \frac{\kappa^{2}\mu}{2w^{2}}\varepsilon^{ijk}R_{il}\nabla_{j}R^{l}_{k} - \frac{\kappa^{2}\mu^{2}}{8}R_{ij}R^{ij}
  \\ &+ \frac{\kappa^{2}\mu^{2}}{8(1-3\lambda)} \left(\frac{1-4\lambda}{4}R^{2} + \Lambda_{W}R - 3\Lambda_{W}^{2}\right)\bigg\},
\end{align}
where $\mu$, $w$, and $\Lambda_{W}$ are constant parameters and $C_{ij}$ is the Cotton tensor defined by
\begin{equation}\label{CottonTensor}
  C^{ij} \equiv \varepsilon^{ikl}\nabla_{k} \left(R^{j}_{~l}-\frac{1}{4}R\delta^{j}_{~l}\right).
\end{equation}
Comparing with the Einstein-Hilbert action in the ADM formalism, the speed of light, Newton's constant and the cosmological constant emerge as
\begin{equation}\label{cGL}
  c = \frac{\kappa^{2}\mu}{4} \sqrt{\frac{\Lambda_{W}}{1-3\lambda}}, \quad G_{N} = \frac{\kappa^{2}}{32\pi c}, \quad \Lambda = \frac{3}{2} \Lambda_{W}.
\end{equation}
Furthermore the requirement that this action be equivalent to the standard Einstein-Hilbert action in the IR limit requires that the running constant $\lambda$ takes its relativistic value $\lambda = 1$. It is also important to note from the former expressions that for $\lambda > \tfrac{1}{3}$ we must have a negative cosmological constant incompatible with observations. However, as pointed out in \cite{Pope}, this problem can be solved by taking an analytical continuation of the constants $\mu$ and $w^{2}$, namely $\mu \rightarrow i\mu$ and $w^{2} \rightarrow -iw^{2}$. This changes the sign of the potential term in the action \eqref{HoravaAction}, leaving the kinetic term intact and with emergent speed of light $c = \tfrac{1}{4}\kappa^{2}\mu \sqrt{\Lambda_{W}/(3\lambda-1)}$. In this case the cosmological constant is, as desired, positive for $\lambda > \tfrac{1}{3}$.

\subsection{Hamiltonian formulation}\label{HamiltonianFormulation}
As already pointed out in \cite{Horava1} a Hamiltonian formulation for gravity with anisotropic scaling results particularly natural because of the $3+1$ split of spacetime. In this sense the metric fields become canonical variables and the Hamiltonian for the theory may be written, as in GR, as a sum of constraints, namely,
\begin{equation}
  H = \int d^{D}\mathbf{x} \left( N\mathcal{H}_{\bot} + N^{i}\mathcal{H}_{i} \right),
\end{equation}
with $\mathcal{H}_{\bot}$ and $\mathcal{H}_{i}$ being the Hamiltonian and momentum constraints, respectively, given by
\begin{eqnarray}
  \nonumber \mathcal{H}_{\bot} &=& \frac{\kappa^{2}}{2\sqrt{g}} \Pi^{ij}\mathcal{G}_{ijkl}\Pi^{kl} + \frac{2\sqrt{g}}{\kappa^{2}}\mathcal{V},
  \\ \mathcal{H}^{i} &=& -2\nabla_{j}\Pi^{ij},
\end{eqnarray}
where
\begin{equation}
  \Pi^{ij}= \frac{\delta S}{\delta \dot{g}_{ij}} = \frac{2\sqrt{g}}{\kappa^{2}} G^{ijkl}K_{kl},
\end{equation}
are the canonical momenta conjugate to the spatial metric and
\begin{equation}
  G^{ijkl} = \frac{1}{2}\left(g^{ik}g^{jl} + g^{il}g^{jk}\right) - \lambda g^{ij}g^{kl}
\end{equation}
is the generalized DeWitt metric and $\mathcal{G}_{ijkl}$ its inverse. Note that these Hamiltonian and momentum constraints preserve the same structure of the GR ones in the ADM formalism \cite{Kiefer1}. However, if the projectability condition is satisfied the Hamiltonian constraint becomes nonlocal and it turns out that the constraint algebra of the theory is slightly different from that of GR \cite{Horava1}. In this case one must consider the spatially integrated constraint
\begin{equation}\label{H0}
  \mathcal{H}_{0} \equiv \int d^{D}\mathbf{x} \mathcal{H}_{\bot}.
\end{equation}
Here we may proceed to quantize the model following the Dirac's recipe assuming the canonical variables obey usual commutation relations and imposing the constraints on the state vector. So, the quantum version of the integral constraint $\mathcal{H}_{0}$ together with the momentum constraints $\mathcal{H}_{i}$ will play the role of the Wheeler-DeWitt equations
\begin{equation}
  \hat{\mathcal{H}}_{0}\psi = 0, \qquad \hat{\mathcal{H}}_{i}\psi = 0.
\end{equation}
In Sec. \ref{KSHL:QM} we implement this quantization program to the KS cosmological model within the minimal version of Ho\v{r}ava-Lifshitz gravity described by the action \eqref{HoravaAction}, with the aim to explore the quantum properties of this theory. For this purpose we review how this in done in the context of GR and derive the already known results for the KS model in the next section.

\section{Kantowski-Sachs Model in General Relativity}\label{KSGR}
\subsection{The quantum model}\label{KSGR:QM}
Let us then start by reviewing the well-known quantum and classical properties of the KS universe, which is one of the simplest homogenous and anisotropic models, in the context of GR. The metric for this model is
{\small
\begin{align}\label{KSsolution}
  \nonumber ds^{2} =& - \left(\frac{\Lambda t^{2}}{3} + \frac{2m}{t} - 1 \right)^{-1} dt^{2} + \left(\frac{\Lambda t^{2}}{3} + \frac{2m}{t} - 1 \right)dr^{2}
  \\& + t^{2} (d\theta^{2} + \sin^{2}\theta d\phi^{2}).
\end{align}
}
which is a solution of the Einstein field equations with cosmological constant. A convenient parametrization for this metric due to Misner \cite{Misner1972} is
\begin{align}\label{KSmetric}
  \nonumber ds^2 =& -N^{2}dt^{2} + e^{2\sqrt{3}\beta}dr^{2}
  \\& + e^{-2\sqrt{3}\beta - 2\sqrt{3}\Omega} (d\theta^{2} + \sin^{2}\theta d\phi^{2}),
\end{align}
where $N$ represents the lapse function, while $\Omega$ and $\beta$ parametrize the spatial-metric components of this anisotropic model. In the quantum gravity context one must allow the parameters $\Omega$ and $\beta$ to be completely arbitrary, but in the study of homogeneous universes, the metric depends only on the time parameter. As a consequence, a model with a finite-dimensional configuration space arises known as minisuperspace and whose variables are the three-metric components, parametrized in this case by $\Omega$ and $\beta$ \cite{Kiefer1,Ryan1,Ryan2}.

The minisuperspace quantization of the KS model, following the lines of Sec. \ref{HamiltonianFormulation}, comes through the implementation of the Hamiltonian and momentum constraints on the state vector. This has been carried out in \cite{Misner1972} where the particular parametrization \eqref{KSmetric} was conveniently chosen so that the corresponding WDW equation for this model with a particular factor ordering adopts the form of a Klein-Gordon equation
{\small
\begin{align}\label{WDW:GR}
  \nonumber \bigg\{&-\frac{\partial^{2}}{\partial\Omega^{2}} + \frac{\partial^{2}}{\partial\beta^{2}} + 48e^{-2\sqrt{3}\Omega}\left[1 - \Lambda e^{-2\sqrt{3}(\beta+\Omega)} \right]\bigg\} \psi(\Omega,\beta)
  \\& = 0.
\end{align}
}
For this model the momentum constraints are satisfied identically \cite{Ryan1,Ryan2}. So, the analysis of this model can be performed in a similar manner as in standard quantum mechanics with $\Omega$ and $\beta$ being the minisuperspace ``coordinates'' chosen to describe the KS quantum cosmological model. The solution to this equation with $\Lambda=0$ is given by
\begin{equation}\label{WDW:GR:sol}
  \psi_{\nu}^{\pm}(\Omega, \beta) = e^{\pm i \nu \sqrt{3} \beta} K_{i\nu}(4e^{-\sqrt{3}\Omega}),
\end{equation}
where $\nu$ is a separation constant and $K_{i\nu}$ are the modified Bessel functions (Macdonald functions) of imaginary order. Although this wave function is, in general, not normalizable, we may analyze its physical properties by constructing a wave packet \cite{Kiefer2}. This is done in Sec. \ref{KSHL:QM} in order to analyze the quantum solutions obtained in Ho\v{r}ava-Lifshitz gravity and to compare them with the ones derived in GR.

\subsection{The semiclassical model (WKB approach)}\label{KSGR:WKB}
Now in order to obtain the (semi)classical analog of the WDW equation \eqref{WDW:GR} we proceed to apply the Wentzel-Kramers-Brillouin (WKB) method \cite{Kiefer2}. For this we assume separability of the wave function in its arguments $\beta$ and $\Omega$ and propose the ansatz
\begin{equation}
  \psi(\Omega, \beta) = e^{i[S_{1}(\Omega) + S_{2}(\beta)]}.
\end{equation}
The WKB approximation is reached in the limit
\begin{align}
  \nonumber &\left|\frac{\partial^{2}S_{1}(\Omega)}{\partial\Omega^{2}}\right| \ll \left(\frac{\partial S_{1}(\Omega)}{\partial\Omega}\right)^{2},
  \\& \left|\frac{\partial^{2}S_{2}(\beta)}{\partial\beta^{2}}\right| \ll \left(\frac{\partial S_{2}(\beta)}{\partial\beta}\right)^{2}.
\end{align}
Then substituting this form of the wave function back into \eqref{WDW:GR} together with the conditions on the derivatives of the $S$-functions we obtain the Einstein-Hamilton-Jacobi equation
\begin{align}
  \nonumber -&\left(\frac{dS_{1}(\Omega)}{d\Omega}\right)^{2} + \left(\frac{dS_{2}(\beta)}{d\beta}\right)^{2}
  \\&- 48e^{-2\sqrt{3}\Omega} \left[1 - \Lambda e^{-2\sqrt{3}(\beta+\Omega)} \right] = 0.
\end{align}
Now, we identify $\frac{dS_{1}(\Omega)}{d\Omega} \rightarrow \Pi_{\Omega}$ and $\frac{dS_{2}(\beta)}{d\beta} \rightarrow \Pi_{\beta}$, where
\begin{align}\label{momenta:GR}
  \nonumber &\Pi_{\Omega} = -\frac{12}{N}e^{-\sqrt{3}(\beta+2\Omega)}\dot{\Omega}\quad \textrm{and}
  \\& \Pi_{\beta} = \frac{12}{N}e^{-\sqrt{3}(\beta+2\Omega)}\dot{\beta}.
\end{align}
These standard identifications in the WKB procedure lead us to obtain the classical equation
\begin{equation}
  \frac{3}{N^{2}}(\dot\Omega^{2} - \dot\beta^{2}) + e^{2\sqrt{3}(\beta+\Omega)} \left[1 - \Lambda e^{-2\sqrt{3}(\beta+\Omega)} \right]=0.
\end{equation}
It is only up to this classical level that we can identify $\Omega$ and $\beta$ as functions of time. Then, making use of the Misner parametrization \eqref{KSmetric}, taking $e^{-2\sqrt{3}\beta-2\sqrt{3}\Omega}=t^{2}$ and identifying $N^{2}=e^{-2\sqrt{3}\beta}$, we get the equation
\begin{equation}\label{ClassicalEquation:Einstein}
  e^{-2\sqrt{3}\Omega}(1+2\sqrt{3}t\dot{\Omega}) - t^{2} \left(1 - \Lambda t^{2}\right)=0,
\end{equation}
whose solution is
\begin{equation}\label{ClassicalSolution:Einstein}
  e^{-2\sqrt{3}\Omega} = \frac{\Lambda t^{4}}{3} + k t - t^{2},
\end{equation}
where $k$ is an integration constant. This brings us back to the metric \eqref{KSsolution} if we take $k=2m$. So as expected the WKB method applied to the WDW equation \eqref{WDW:GR} gives the classical equation \eqref{ClassicalEquation:Einstein}, and its solution \eqref{ClassicalSolution:Einstein} is the same as the one obtained through the classical Einstein's field equations.

\begin{figure*}[htb!]
  \centering
  \subfigure[]{\includegraphics[width=0.4\textwidth]{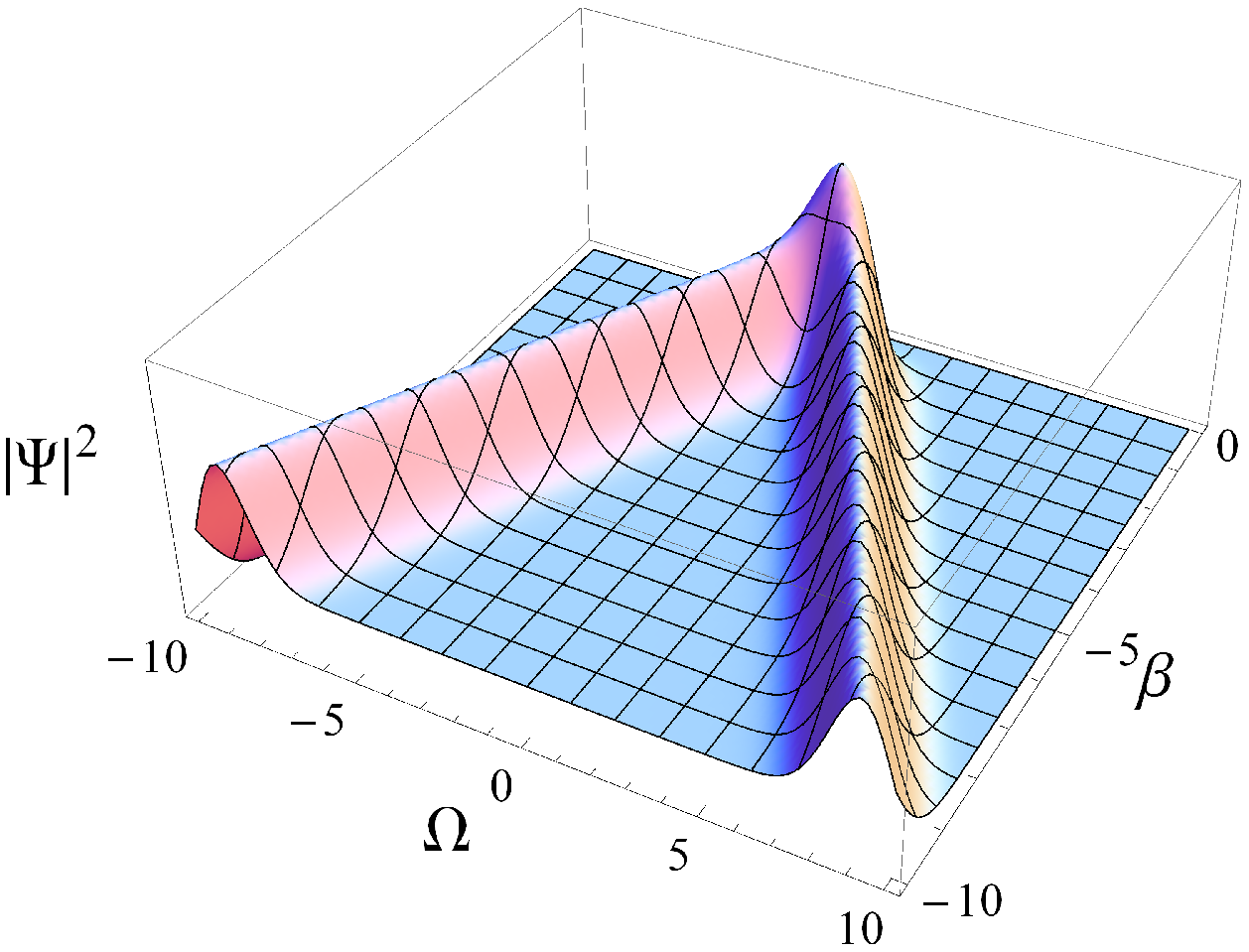}\label{WPHorava}}\qquad \quad
  \subfigure[]{\includegraphics[width=0.4\textwidth]{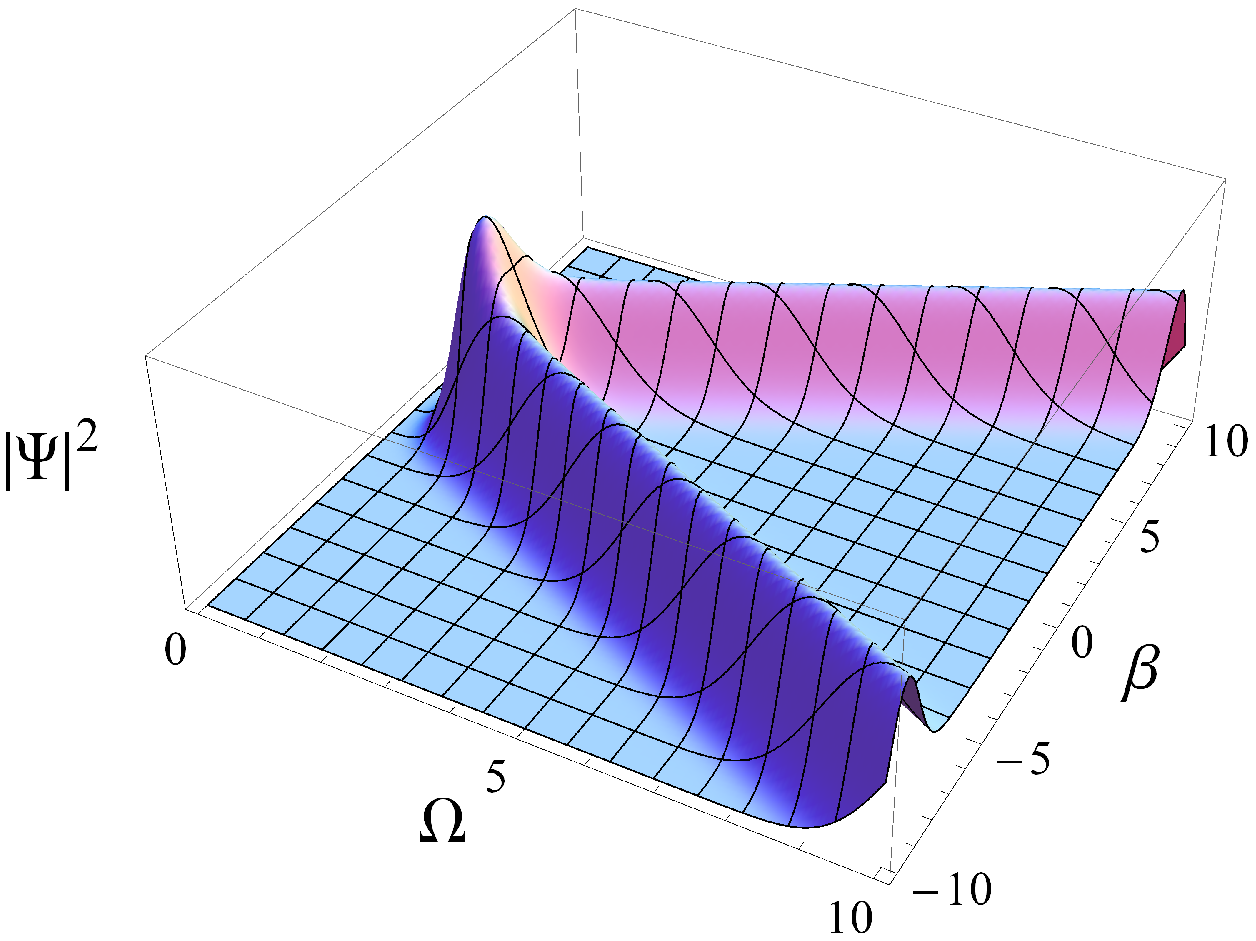}\label{WPEinstein}}
  \label{WPHL:subfigures1}
  \caption{(color online). Variation of $|\Psi|^{2}$ with respect to $\Omega$ and $\beta$: (a) case $\lambda = 1$ in Ho\v{r}ava-Lifshitz gravity, (b) the GR case (IR limit).}
\end{figure*}

\section{Kantowski-Sachs Model in Ho\v{r}ava-Lifshitz gravity}\label{KSHL}
\subsection{The quantum model}\label{KSHL:QM}
Now let us proceed similarly to study the KS model within the framework of the minimal theory proposed by Ho\v{r}ava. We consider here the original action \eqref{HoravaAction} as well as its analytical continuation described in Sec. \ref{TheAction}. In this case, quantization comes through the implementation of the spatially integrated Hamiltonian constraint \eqref{H0} on the state vector. So, the resulting WDW equations are
\begin{align}\label{WDW:HL}
  \nonumber \bigg\{ & \frac{1}{2}(\lambda-3)\frac{\partial^{2}}{\partial\Omega^{2}} - 2(\lambda-1)\frac{\partial}{\partial\Omega}\frac{\partial}{\partial\beta}  +(2\lambda-1)\frac{\partial^{2}}{\partial\beta^{2}}
  \\ \nonumber &\mp 3\mu^{2}\Lambda_{W}e^{-2\sqrt{3}\Omega} \bigg[2 - 3\Lambda_{W}e^{-2\sqrt{3}(\beta+\Omega)}
  \\ &+ \frac{(2\lambda-1)}{\Lambda_{W}}e^{2\sqrt{3}(\beta+\Omega)}\bigg]\bigg\}\psi_{1,2}(\Omega,\beta)=0,
\end{align}
where $\psi_{1}$ and $\psi_{2}$ stand for the wave functions associated with the $(-)$ and $(+)$ signs in the potential corresponding to the original and the analytically continued actions respectively. These quantum equations considerably differ from the usual WDW equation \eqref{WDW:GR} but notice that for this particular model with $\lambda=1$ and making use of the expressions \eqref{cGL} in units such that $c=1$ and $16\pi G_{N} = 1$, they reduce to \eqref{WDW:GR} up to the first two terms of the potential in \eqref{WDW:HL}. So the IR limit gives the same behavior obtained in GR. However, there is an additional term in the potential coming from the higher-order terms in \eqref{HoravaAction} and which is responsible for the UV behavior of the model. Thus, since an analytical solution for this pair of equations is difficult to be found let us consider the UV limit where this last term in the potential dominates. The reduced WDW equations are
\begin{align}\label{WDW:HL:UV}
   \nonumber \bigg[ &\frac{1}{2}(\lambda-3)\frac{\partial^{2}}{\partial\Omega^{2}} - 2(\lambda-1)\frac{\partial}{\partial\Omega}\frac{\partial}{\partial\beta} + (2\lambda-1)\frac{\partial^{2}}{\partial\beta^{2}}
   \\ & \mp 3\mu^{2}(2\lambda-1)e^{2\sqrt{3}\beta}\bigg]\psi_{1,2}(\Omega, \beta) = 0.
\end{align}
The solutions to these equations are given in terms of Macdonald and Bessel functions, respectively
{\small
\begin{equation}\label{WDW:HL:UV:sol:normalized}
  \begin{bmatrix}
    \psi_{1\nu} \\
    \psi_{2\nu}
  \end{bmatrix}
  = e^{\pm i\nu\sqrt{3}\Omega} (\mu e^{\sqrt{3}\beta})^{\pm (\frac{\lambda-1}{2\lambda-1})i\nu}  \begin{bmatrix}
    K_{\frac{\sqrt{3\lambda-1}}{\sqrt{2}(2\lambda-1)}i\nu}(\mu e^{\sqrt{3}\beta})
    \\ J_{\frac{\sqrt{3\lambda-1}}{\sqrt{2}(2\lambda-1)}i\nu}(\mu e^{\sqrt{3}\beta})
  \end{bmatrix},
\end{equation}
}
with $\nu$ a separation constant. Despite the bulky form of this solution, $\psi_{1\nu}$ takes a familiar structure for the particular case $\lambda=1$ where the wave function is
\begin{equation}\label{WDW:HL:UV:sol:lambda1}
  \psi_{1\nu}(\Omega,\beta)= e^{\pm i\nu\sqrt{3}\Omega} K_{i\nu}(\mu e^{\sqrt{3}\beta}).
\end{equation}
This particular solution resembles the one obtained in GR \eqref{WDW:GR:sol}, but it results that the roles of the minisuperspace variables are now interchanged. So, a simple but completely different physical result arises. It is noteworthy here that for some of the results in this paper we have considered that the free parameter $\lambda$ takes the specific value $\lambda = 1$. However, $\lambda$ represents a dynamical coupling constant which could run under the renormalization group flow since it is not protected by any symmetry. Then it is expected that in the UV regime $\lambda$ runs away from its desired IR fixed point $\lambda = 1$, where GR should be recovered \cite{Horava2,Mukohyama,Padilla}. However, using this value is not problematic and can be used for simplicity with the purpose of analyzing the quantum properties of the model.

\begin{figure*}[t!]
  \centering
  \subfigure[]{\includegraphics[width=0.4\textwidth]{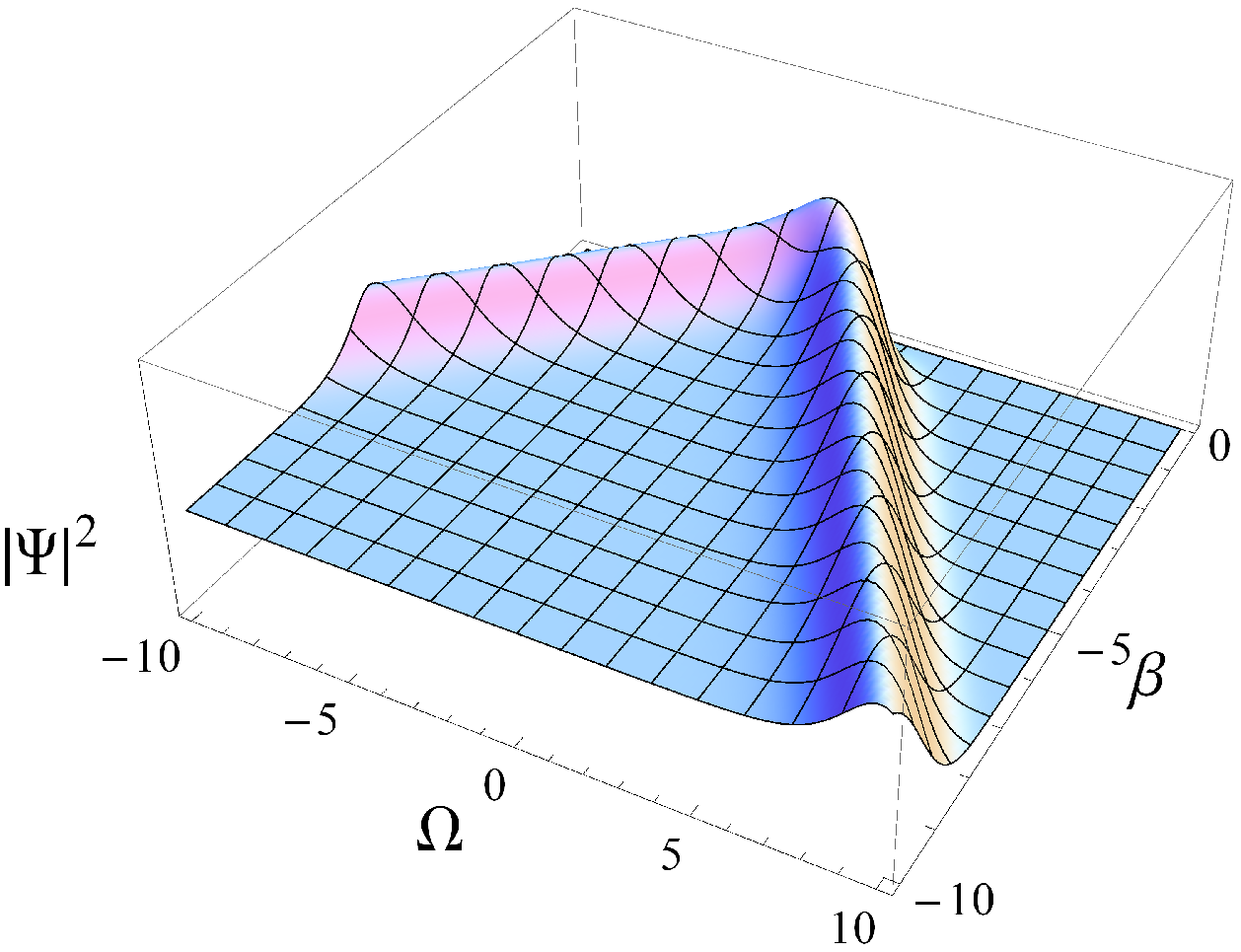}\label{WPHorava075}}\qquad \quad
  \subfigure[]{\includegraphics[width=0.4\textwidth]{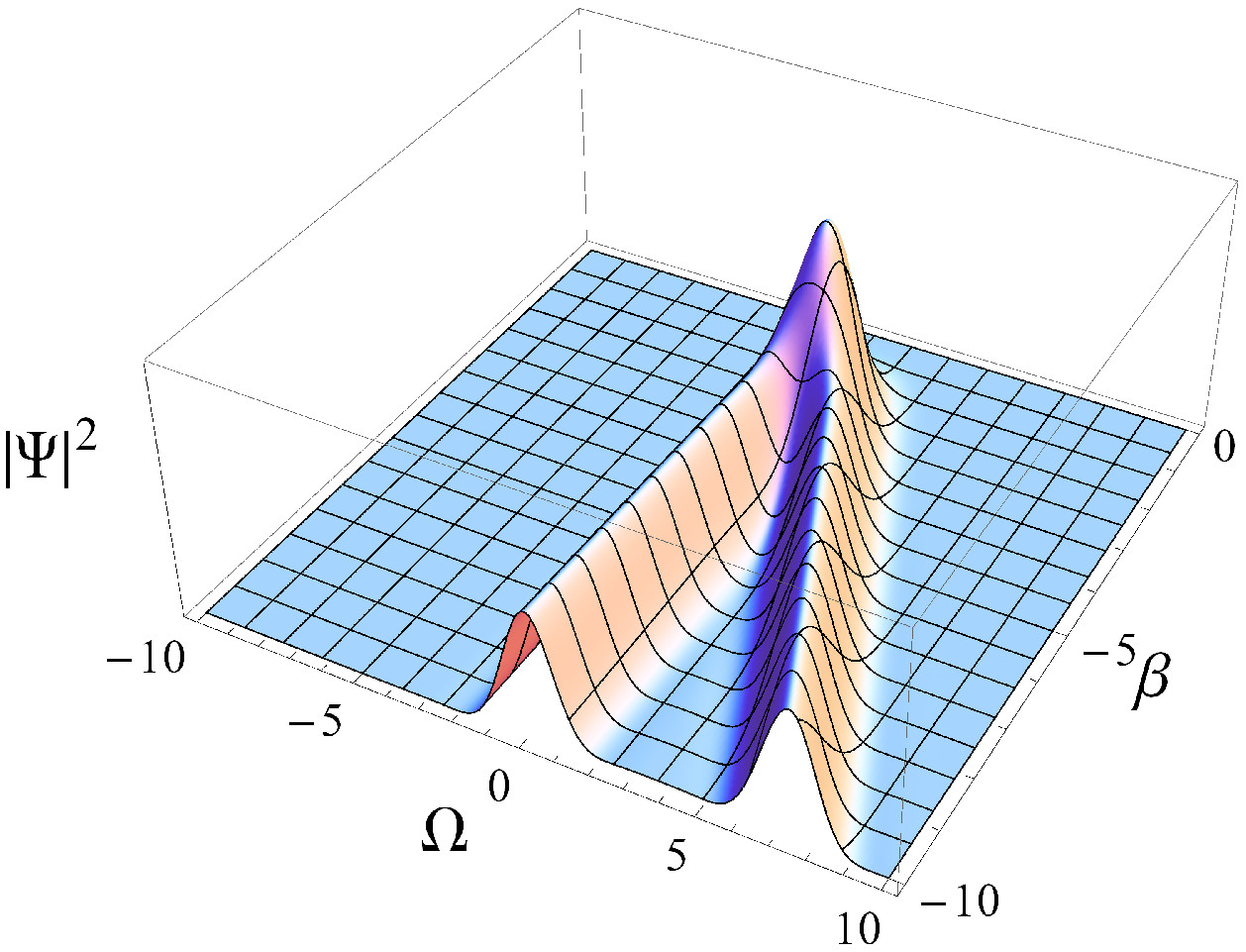}\label{WPHorava3}}
  \label{WPHL:subfigures2}
  \caption{(color online). Variation of $|\Psi|^{2}$ with respect to $\Omega$ and $\beta$: (a) case $\lambda = 0.75$, (b) case $\lambda = 3$.}
\end{figure*}

In order to see the consequences of the UV corrections more clearly, a general solution may be obtained by superposition of the quantum cosmological solutions of the WDW equation with an appropriate amplitude in the sense of the principle of constructive interference \cite{Kiefer1,Kiefer2}. Let us here consider a wave packet weighted by a Gaussian centered in $\nu = \bar\nu$ and with standard deviation $\sigma$ for $\psi_{1\nu}$
\begin{equation}\label{WavePacket}
  \Psi(\Omega,\beta) = \mathcal{N} \int_{-\infty}^{\infty} e^{-\frac{1}{2\sigma^{2}}(\nu - \bar\nu)^{2}} \psi_{1\nu}(\Omega,\beta)d\nu.
\end{equation}
Here $\mathcal{N}$ is a normalization constant. This integral is performed numerically for the specific values $\sigma^{2} = 1/3$ and $\bar\nu=1.3$ with $\mu = 4$, and for different values of $\lambda$. Not only are we interested to see the influence of the higher-order spatial curvature terms in the action, but also we want to study the behavior of the probability depending on the values of the $\beta$ and $\Omega$ variables. Figure \ref{WPHorava} shows the variation of the square of the wave packet magnitude $|\Psi|^{2}$ as a function of the minisuperspace variables $\beta$ and $\Omega$ for the case $\lambda=1$ in the UV region. It can be seen that in this case there is only one absolute maximum representing the most probable state of the Universe around $\beta = -1.5$ and $\Omega = 0$. In contrast, Fig. \ref{WPEinstein} shows the already known IR behavior, the GR one \cite{Obregon}, described by the corresponding solution \eqref{WDW:GR:sol}. Here we also have one stable state for the Universe but around $\beta = 0$ and $\Omega = 1.5$. It can be seen from both graphs that the probability distribution changes drastically, allowing a very different quantum universe. In some sense they are opposed since the minisuperspace variables switch their role between these two limits, UV and IR. This, in fact, can be easily observed by comparing the WDW equation \eqref{WDW:GR} for $\Lambda = 0$ with \eqref{WDW:HL:UV} for $\lambda =1$ under the replacements $\beta \rightarrow \Omega$ and $-\Omega \rightarrow \beta$. Additionally, Figs. \ref{WPHorava075} and \ref{WPHorava3} show the corresponding wave packets for the specific values of $\lambda = 0.75$ and $\lambda = 3$ in the UV region, respectively. It can be seen that although we vary the value of $\lambda$, a single preferred state of the Universe remains in this UV region, but the behavior of the Universe is completely different compared with the GR one. In general, we can see that the Universe would live in a very different quantum state when its behavior is dominated by the higher-order terms in \eqref{HoravaAction}.

Even though this particular exchange in the behavior between the minisuperspace variables $\Omega$ and $\beta$ in the IR and UV regions for $\lambda = 1$ happens for this specific model, this result seems to indicate that for other minisuperspaces and even for more general models, one should expect significant quantum physical differences among the IR and UV regions.

It is worth to mention here that the minisuperspace construction is a procedure to define quantum cosmological models in the search to describe the quantum behavior of the Universe at its very early stages. By defining these models, one necessarily freezes out degrees of freedom, so that these are only simple and probably approximate models of quantum gravity. Nevertheless, this procedure has allowed us to construct a simple quantum cosmological model in the framework of Ho\v{r}ava-Lifshitz gravity and to perform a study of the early Universe analyzing the properties of the corresponding quantum solutions. In this context, it would be of great interest to analyze any imprint of the UV corrections of the theory at the classical level. With this purpose, in the next section we find the temporal evolution by carrying a semiclassical approximation.

\begin{figure*}
  \centering
  \subfigure[]{\includegraphics[width=0.45\textwidth]{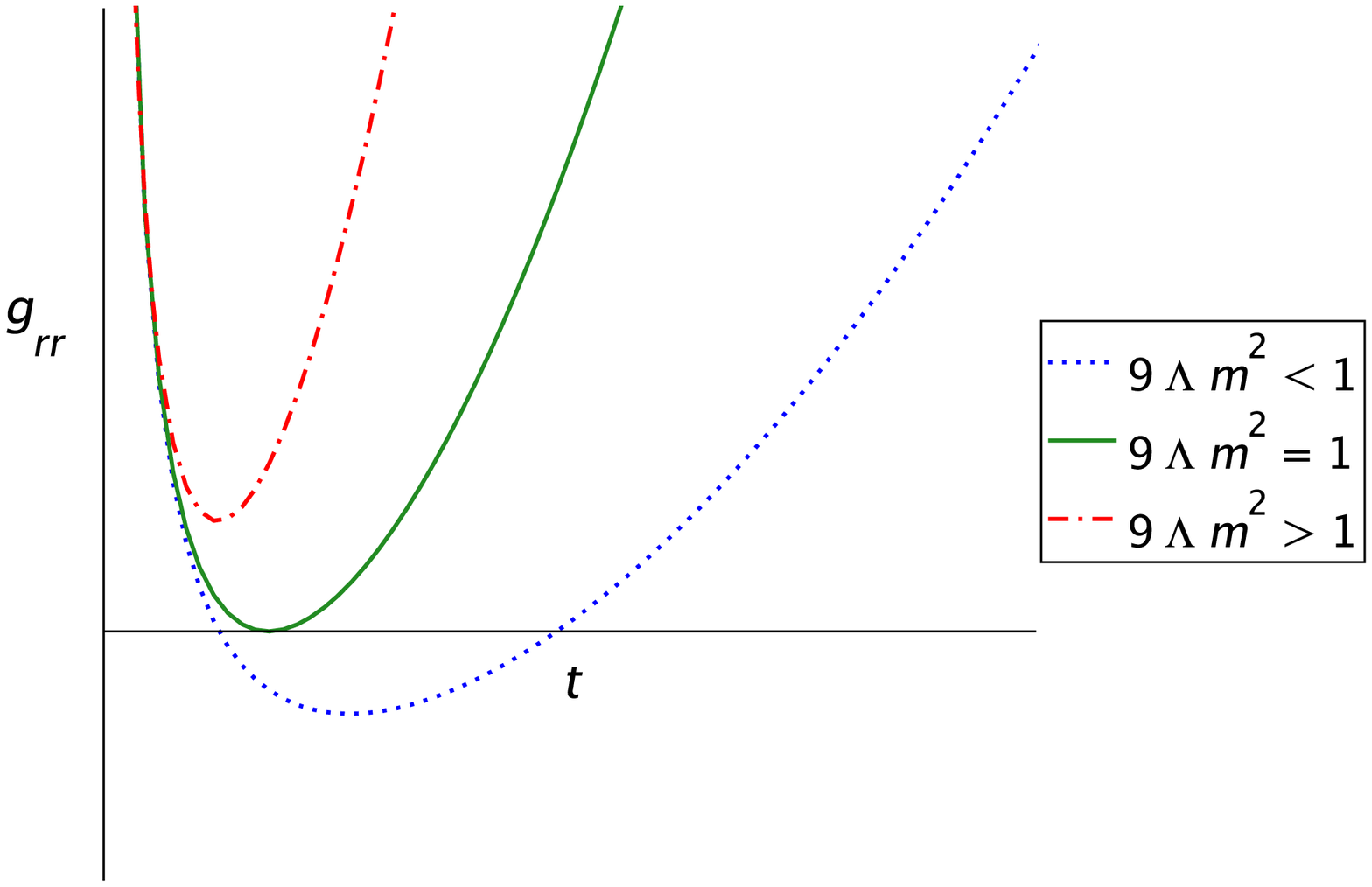}\label{GR:subfig1}}
  \subfigure[]{\includegraphics[width=0.45\textwidth]{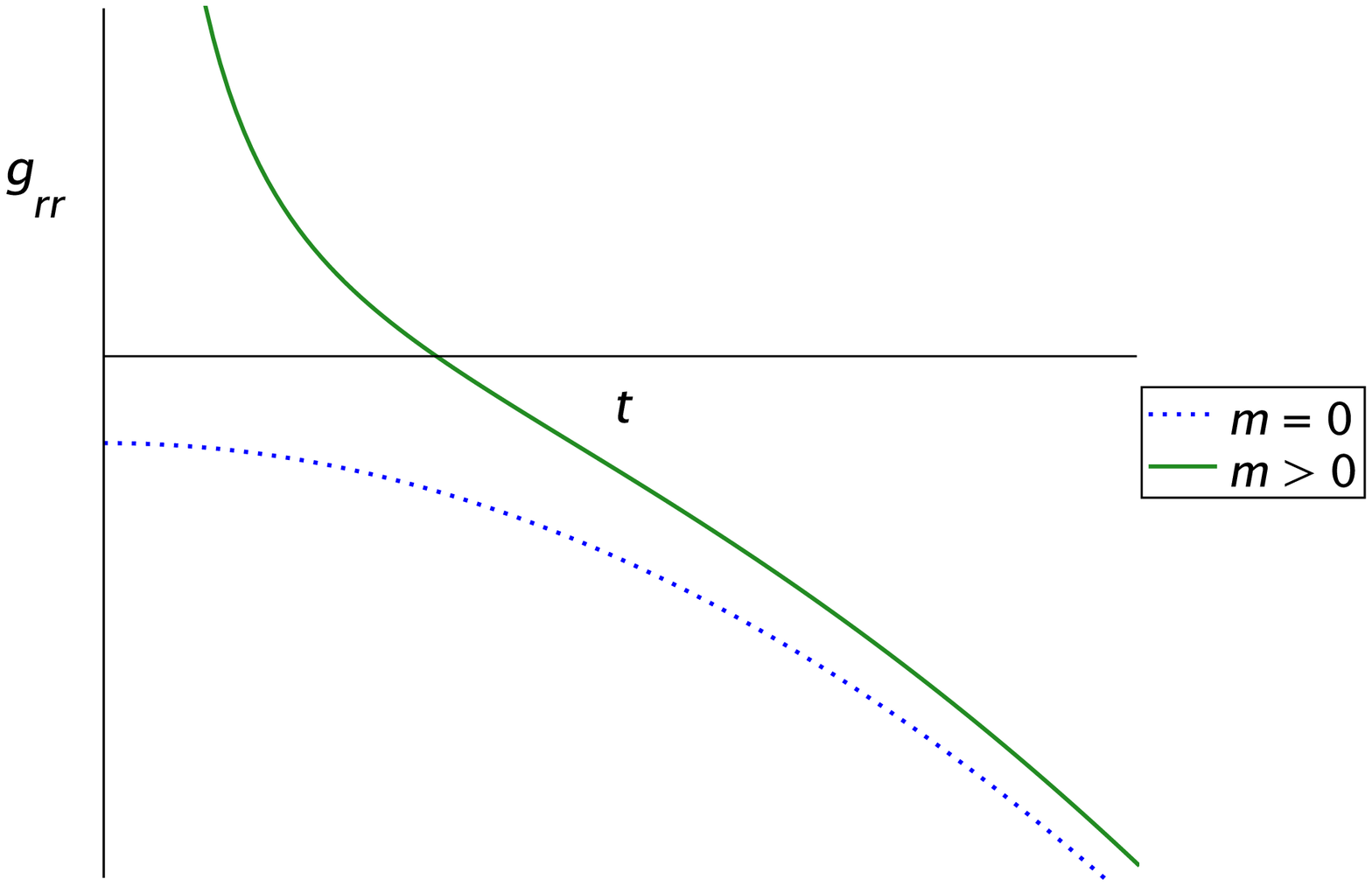}\label{GR:subfig2}}
  \label{GR:subfigures}
  \caption{(color online). Metric coefficient $g_{rr}$ for the KS metric in GR. The event horizons coincide with the zeros of this function. (a) Case $\Lambda > 0$ (b) Case $\Lambda < 0$.}
\end{figure*}

\begin{figure*}
  \centering
  \subfigure[]{\includegraphics[width=0.45\textwidth]{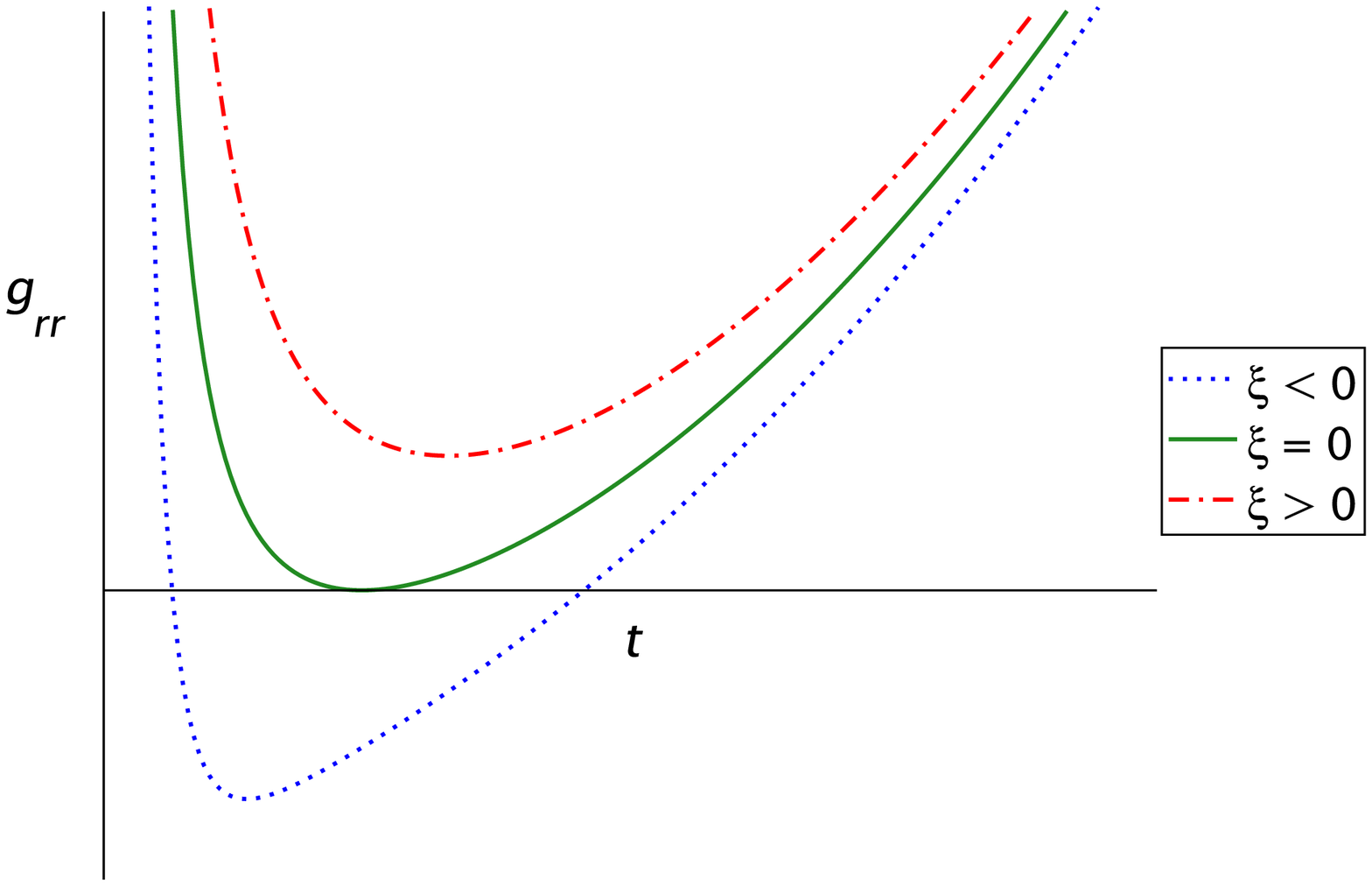}\label{HL:subfig1}}
  \subfigure[]{\includegraphics[width=0.45\textwidth]{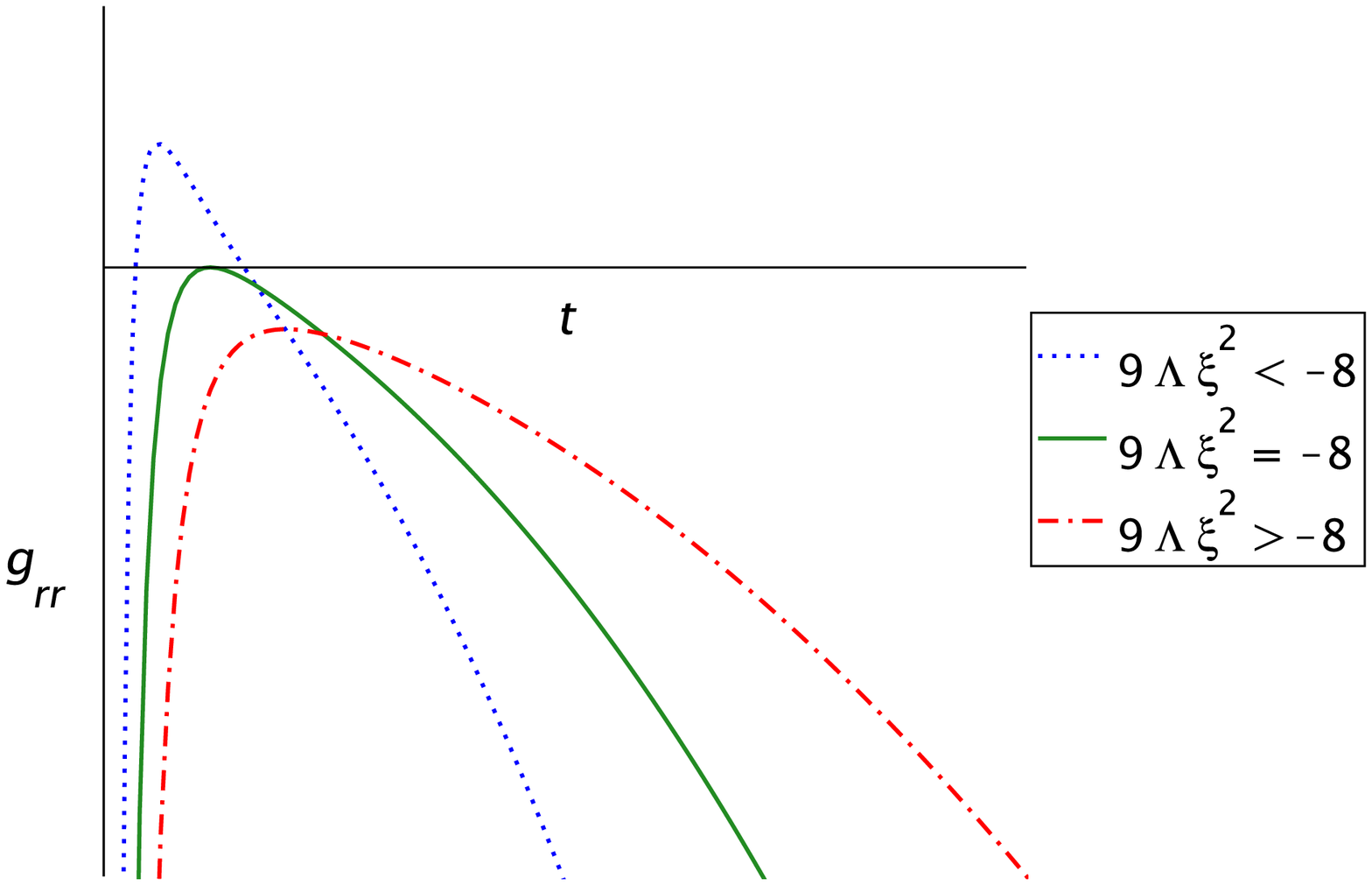}\label{HL:subfig2}}
  \label{HL:subfigures}
  \caption{(color online). Metric coefficient $g_{rr}$ for the KS metric in Ho\v{r}ava-Lifshitz gravity. The event horizons coincide with the zeros of this function. (a) Case $\Lambda > 0$ corresponding to the analytical continuation of the Ho\v{r}ava action, (b) case $\Lambda < 0$ corresponding to the original Ho\v{r}ava action \eqref{HoravaAction}.}
\end{figure*}

\subsection{The semiclassical model (WKB approach)}\label{KSHL:WKB}
For the general WDW equation \eqref{WDW:HL} the standard WKB procedure described in Sec. \ref{KSGR:WKB} results in the following pair of classical equations:
\begin{align}\label{ClassicalEquation:Horava}
  \nonumber &\frac{3 e^{-2\sqrt{3}(\beta +\Omega)}}{N^{2}} \left[(3-\lambda)\dot{\beta}^{2} + 4(1-\lambda)\dot{\beta}\dot{\Omega} - 2(2\lambda-1)\dot{\Omega}^{2}\right]
  \\ \nonumber &\mp \frac{\kappa^{4}\mu^{2}\Lambda_{W}}{16(1-3\lambda)}
  \bigg[2 - 3\Lambda_{W}e^{-2\sqrt{3}(\beta+\Omega)}
  \\ &+ \frac{(2\lambda-1)}{\Lambda_{W}}e^{2\sqrt{3}(\beta+\Omega)}\bigg] = 0.
\end{align}
Now, inspired by the usual KS solution in GR, ma\-king use of the Misner parametrization \eqref{KSmetric}, taking $e^{-2\sqrt{3}\beta-2\sqrt{3}\Omega}=t^{2}$ and identifying $N^{2}=e^{-2\sqrt{3}\beta}$ we get the equation
\begin{align}
  \nonumber &3(1-\lambda)t^{2}\dot{\Omega}^{2} + 2(1+\lambda)\sqrt{3}t\dot{\Omega} + (3-\lambda)
  \\ &- c_{\pm}^{2}t^{2}e^{2\sqrt{3}\Omega} \left[2 - 3\Lambda_{W}t^{2} + \frac{(2\lambda-1)}{\Lambda_{W}t^{2}}\right] = 0,
\end{align}
where $c_{-}$ and $c_{+}$ represent the speed of light for the original and the analytically continued actions, respectively. An analytical solution for this equation is difficult to be found so let us consider the case $\lambda = 1$. The solution for this particular case, making use of the expressions \eqref{cGL} in units such that $c_{\pm} = 1$, is
\begin{equation}
  e^{-2\sqrt{3}\Omega} = \frac{\Lambda t^{4}}{3} + 2\xi t+ \frac{3}{4\Lambda} - t^{2},
\end{equation}
where $\xi$ is an integration constant \footnote{This constant is usually assumed to be positive in the context of GR, where it is identified with a mass parameter for a black hole immerse in an asymptotically (anti)-de Sitter space due to the well-known diffeomorphism between the KS metric and the Schwarzschild (anti)-de Sitter spacetime \cite{Zacarias2}.}. Then we have the metric
\begin{align}\label{KSsolutionHorava}
  \nonumber ds^{2} =&- \left( \frac{\Lambda t^{2}}{3} + \frac{2\xi}{t} + \frac{3}{4\Lambda t^{2}} - 1 \right)^{-1} dt^{2}
  \\ \nonumber &+ \left(\frac{\Lambda t^{2}}{3} + \frac{2\xi}{t} + \frac{3}{4\Lambda t^{2}} - 1 \right)dr^{2}
  \\ &+ t^{2} (d\theta^{2} + \sin^{2}\theta d\varphi^{2}),
\end{align}
which is a solution of the original action with $\Lambda < 0$ and of its analytical continuation with $\Lambda > 0$. This solution is a generalization of the KS metric \eqref{KSsolution} with an additional term proportional to $1/t^{2}$ coming from the second-order spatial curvature terms in the action \eqref{HoravaAction} and which becomes more significant for very small values of $t$ as it would be expected. In order to understand the consequences of this extra term let us study the properties of this generalized KS metric by comparing it with the GR solution \eqref{KSsolution}.

It is well known that the KS metric in GR \eqref{KSsolution} with $\Lambda > 0$ and $9\Lambda m^{2} < 1$ shows two event horizons, a black hole horizon located at $t = t_{h}$ and a cosmological horizon in $t = t_{c}$. So the spacetime is dynamic at $t < t_{h}$ and $t > t_{c}$, and static in the region $t_{h} < t < t_{c}$. If $9\Lambda m^{2} = 1$ both horizons coincide at $t = 3m$, and if $9\Lambda m^{2} > 1$ the spacetime is dynamic at all $t > 0$. For the case $\Lambda < 0$ and $m>0$ there is only one horizon located at $t_{+}$. Here the spacetime is dynamic at $t < t_{+}$ and static at $t > t_{+}$. For this case if $m = 0$ no horizon will be present. This behavior is depicted in Figs. \ref{GR:subfig1} and \ref{GR:subfig2} where the horizons coincide with the zeros of the $g_{rr}$ metric coefficient of \eqref{KSsolution}. This metric is, in general, singular at $t = 0$ \cite{Stuchlik}.

Now, for the generalized KS metric \eqref{KSsolutionHorava} the structure of the spacetime changes completely. In the case $\Lambda > 0$, corresponding to the analytical continuation of the action \eqref{HoravaAction}, we have that
\begin{itemize}
  \item if $\xi < 0$ there are two singular points located at $t_{+}$ and $t_{++}$. The spacetime is dynamic at $t < t_{+}$ and $t > t_{++}$.
  \item if $\xi = 0$ both singularities coincide at $t = \sqrt{\frac{3}{2\Lambda}}$.
  \item if $\xi > 0$ the spacetime is dynamic at all $t > 0$.
\end{itemize}
Now, for the case $\Lambda < 0$, corresponding to the original action \eqref{HoravaAction},
\begin{itemize}
  \item if $9 \Lambda \xi^{2} < -8$ and $\xi > 0$ there are two singular points located at $t_{-}$ and $t_{--}$. The spacetime is dynamic in the region $t_{-} < t < t_{--}$.
  \item if $9 \Lambda \xi^{2} = -8$ and $\xi > 0$ both singularities coincide at $t = \frac{3\xi}{4}$. The metric coefficient is negative elsewhere, resulting in an unphysical solution.
  \item if $9 \Lambda \xi^{2} > -8$ the metric coefficient is negative for all $t$,x| giving an unphysical solution.
\end{itemize}
The previous behavior can be observed in Figs. \ref{HL:subfig1} and \ref{HL:subfig2} where the singular points of this generalized KS metric coincide with the zeros of the $g_{rr}$ metric coefficient of \eqref{KSsolutionHorava}. In both cases this generalized metric is also singular at $t = 0$.

In order to exhibit the essential singularities of this metric one may go to the concept of invariants built on the curvature tensor. A very common and useful choice is the Kretschmann invariant $R_{\mu\nu\alpha\beta}R^{\mu\nu\alpha\beta}$. However, we must remember that in Ho\v{r}ava-Lifshitz gravity we do not have the full set of diffeomorphisms present in GR. For this reason, one must consider scalars invariant under $\text{Diff}_{\mathcal{F}}(\mathcal{M})$ constructed from the extrinsic curvature tensor $K_{ij}$, the three-dimensional Riemann tensor and their derivatives \cite{Cai3}. In this case, we consider the Kretschmann invariant built on the spatial curvature tensor and the three quantity $K = g^{ij}K_{ij}$, which for this metric are
\begin{equation}
  R_{ijkl}R^{ijkl} =\frac{4}{t^{4}}, \quad K = \frac{4\Lambda^{2}t^{4} - 8\Lambda t^{2} + 12\xi\Lambda t + 3}{4\Lambda t^{3} \left( \frac{\Lambda t^{2}}{3} + \frac{2\xi}{t} + \frac{3}{4\Lambda t^{2}} - 1 \right)^{1/2}}.
\end{equation}
Clearly it can be seen that according to the Kretschmann invariant the metric \eqref{KSsolutionHorava} is singular at $t = 0$ as it happens in GR. However, the scalar $K$ becomes singular also at all the points previously listed dictated by the zeros of the $g_{rr}$ metric coefficient in \eqref{KSsolutionHorava}. These are scalar singularities that cannot be removed by the restricted coordinate transformations \eqref{symmetries} and which are completely absent in GR.

\section{Discussion and Conclusions}\label{Conclusions}
In this work we have constructed a simple quantum model into the minimal version of Ho\v{r}ava-Lifshitz gravity. We chose the KS cosmological model for this purpose. A WDW equation was derived in this framework and quantum cosmological solutions for the UV limit of this model were analytically obtained. The analysis of the resulting solutions was performed by constructing wave packets weighted by a Gaussian amplitude and plotting the probability distribution for different values of the parameter $\lambda$. Then, these results were compared with the already known result derived in GR. It was observed that although we vary the value of $\lambda$, a single preferred state of the Universe seems to remain, however, the behavior of the Universe changes drastically compared with the GR one. This happens since the probability distribution is completely modified allowing the existence of the Universe in a very different quantum state dictated by the domain of the UV-corrections terms in the action. This is clearly illustrated in the simple case $\lambda = 1$, where the minisuperspace variables $\beta$ and $\Omega$ switch their roles between the UV and IR limits. All these observations seem to indicate that due to the contribution of the higher-order terms in the action one should expect very different physical results between the IR and UV regions even for other more general quantum gravity models within the minimal version of Ho\v{r}ava-Lifshitz gravity. In particular, for quantum cosmological models in this framework, a completely different behavior of the Universe would be expected at its very early stages where the anisotropic scaling between space and time would play a significant role.

In this context, we also proceeded to extract dynamical information of the model in order to see the effects of the UV corrections in the classical arena. This was achieved by means of a WKB approximation applied to the complete WDW equation \eqref{WDW:HL}. It is observed that if $\lambda$ takes its relativistic value $\lambda = 1$ a generalized KS metric is obtained, which differs from the usual KS solution in general relativity by an additional term arising from the higher-order curvature terms in the action dominating the behavior of the solution for small values of the time parameter as expected. We have discussed the physical properties of this solution by analyzing the metric coefficients and comparing them with those of the usual KS solution in GR. Particularly, the presence of singular points has been qualitatively described for both signs of the cosmological constant showing significant differences with the KS spacetime structure in GR. By means of the appropriate curvature invariants we have shown that the solution has no horizons but singularities.

\begin{acknowledgments}
  The authors would like to thank S. Zacar\'ias for useful comments on the manuscript. J.A.P acknowledges support from the CONACYT-M\'exico Scholarship Programme. This work was partially supported by CONACYT Grant No. 135023, PROMEP, and UG Projects.
\end{acknowledgments}


%

\end{document}